\documentclass[conference,10pt,twocolumn]{IEEEtran}

\usepackage[dvips]{graphicx}
\usepackage[cmex10]{amsmath}
\usepackage{subfigure}
\usepackage{amsmath}
\usepackage{color}
\usepackage{url}
\usepackage{algpseudocode}
\usepackage{tikz}
\usetikzlibrary{positioning}
\usepackage{caption}
\usepackage{cite}

\title{Directional Communication Enabled by Mobile Parasitic Elements}

\author{\IEEEauthorblockN{Fikadu T. Dagefu, Jeffrey N. Twigg, Chirag R. Rao, Brian M. Sadler}
\IEEEauthorblockA{U.S. Army Research Laboratory, Adelphi, MD 20783\\}
\textsuperscript{}\{fikadu.t.dagefu, jeffrey.n.twigg, chirag.r.rao, brian.m.sadler6\}.civ@mail.mil\\
}

\begin{document}
\maketitle

\begin{abstract}
Mobile communications in complex environments such as mega-cities is a challenging problem that limits the ability to deploy autonomous agents in support of operations.  Building on recent progress in low frequency networking that utilizes miniature antennas to provide persistent connectivity among agents,  we consider the design and collaborative manipulation of a distributed robotic antenna array to provide directional communications that will enable enhanced networking, interference rejection, and collaborative control.  The use of parasitic elements in a Yagi-Uda type array design avoids the need for synchronization and highly accurate position control among the agents. We utilize physics-based simulations to investigate the feasibility of using mobile agents equipped with an excited antenna element along with a set of support nodes having parasitic elements that adaptively configure to enhance radiation in a desired direction. We take into account mobile node pose uncertainty including element position and angular orientation, as well as ground scattering effects. We pursue an optimal design approach for different types of ground electromagnetic characteristics based on a hybrid full-wave propagation simulation and genetic algorithm optimization. We also present experiment with one mobile node and two static elements.  The results demonstrate the ability to achieve directional low frequency communications that is robust to robotic pose error.




\end{abstract}


\section{Introduction}
\label{sec:intro}
Mobile and robotic communications in complex environments is a challenging problem that limits collaborative control and autonomous operations. While there has been substantial progress in developing highly miniature and efficient low-frequency antennas at the low VHF band \cite{Choi2015electrically,oh2013extremely} to penetrate structures in these complex environments, these antennas are generally omni-directional.  However, the overall gain of electrically small antennas (ESAs) is relatively small due to the highly limited physical aperture of the antenna, and smaller antennas inherently tend to have omni-directional patterns. It is naturally of interest to consider antenna arrays to increase gain, control directional preference, and enable interference rejection.  
We envision an array formed through the coordination of robotic agents that collaboratively manipulate the radiation pattern of a parasitic array through mobility to increase signal gain.  There is extensive research in multi-robot manipulation  \cite{williams1993virtual} and transportation \cite{song2002potential,spletzer2001cooperative} of physical objects, robotic  object manipulation in cooperation with humans \cite{mainprice2013human},  and multi-robot communications maintenance \cite{fink2013robust}.  
Here, we consider the feasibility of collaborative manipulation of a distributed robotic antenna array which maintains the high reliability of radio's signal, and adds direction control and increased effective communication range.





Researchers have demonstrated that radios with these ESAs transmitting at 300mW  have a range of 100s of meters and can penetrate obstacles without a decrease in reliability from multi-path fading \cite{Choi2015electrically,oh2013extremely}. In urban environments, this technology would help maintain a persistent connection between mobile agents as they move through buildings.
There is, however, a reduction in bandwidth when decreasing frequency. These ESAs are more efficient and can provide enough bandwidth for streaming video, as well as sharing images and telemetry. We envision mobile agents equipped with multiple radios for different tasks. For mobile agents this technology might serve as an auxiliary channel for time sensitive and small data-size communication. This low VHF channel is especially important for commanding/controlling robotic agents. Robotic agents are teleoperated into dangerous scenarios where it is easy to drive the robot beyond communication range. Physically recovering the robot is also quite dangerous so loss of communication with a robotic agent often means the robot is considered lost as well. In order to achieve greater stand-off distance between agents, we want to increase the effective power or gain of antenna by placing it inside of a parasitic array.


 
Parasitic arrays have been exploited for various applications (e.g. localization, directional communications) to enable directional radiation using multi-element antenna systems. A parasitic array only requires a single driven element, with other passive elements positioned in a way to provide preferential reflection and phasing and thereby yield directionality. A parasitic array is particularly appealing for implementation by a group of robotic unmanned ground vehicles (UGVs) since these robots can autonomously sense (through camera and lidar) each other and optimize their positions within the array without a radio on each robot. In comparison, a conventional distributed array requires tightly synchronized operation among all the agents, as well as highly accurate estimation of the joint positioning and pose of each agent, requirements that significantly complicate distributed array use. An additional consideration that arises with low frequency operation is mutual coupling that occurs between antenna elements that are closely spaced relative to the signal wavelength. These coupling effects, if not accounted for, can significantly degrade array performance \cite{lui2009note,huang2006mutual,liao2012adaptive,kraus2002antennas}.

To our knowledge, there is no prior work on autonomous positioning of parasitic array elements that are physically disconnected (for example, not on a common platform).  In addition, classic parasitic array designs are typically carried out for elevated deployment, and assume the antenna is in free space.  Here, we consider design for ground based application mounted on small robots, and show that the proximity to the ground can seriously degrade existing designs. 

Related work for mobile robotics includes the use of a rotating reflector for localization \cite{graefenstein2009wireless}, and localizing wildlife carrying tracking devices by employing multiple robots to form an array \cite{cliff2015online}.  In this paper, we focus on the case of using a single excited array element (the director), and note that combinations of multiple exciters and passive elements can be employed \cite{zhang2004pattern} \cite{petit2006mems}, which is an interesting extension for further study in our context.


We seek to improve performance which is measured by increasing gain and focused beam direction, while reducing beamwidth and side lobe levels by leveraging mobile parasitic elements for ground based applications. Our contributions are the following:
\begin{itemize}
\item A physics-based simulation framework for mobile parasitic arrays to quantify the effect of uncertainties in position, orientation, and ground dielectric as described in Section \ref{sec:parameteric_study}. 
\item Experiments showing increased gain by adding parasitic elements in Section \ref{sec:experiment}. 
\item A hybrid full-wave propagation simulation and genetic algorithm based optimization of element configuration in order to adapt to the dielectric properties of the ground in Section \ref{sec:optimal_design}.
\end{itemize}

In Section \ref{sec:basics} we briefly describe the basics of parasitic arrays to put Sections \ref{sec:parameteric_study}, \ref{sec:experiment}, and \ref{sec:optimal_design} in context. 
In our last section we summarize the work and describe the potential for multi-agent robotically reconfigurable parasitic arrays to manipulate radiation patterns in ways that can enhance collaborative autonomous agents.


 
 

\section{Basics of Parasitic Arrays}
\label{sec:basics}

\begin{figure}[t]
  \centering
  \includegraphics[width=0.99\columnwidth]{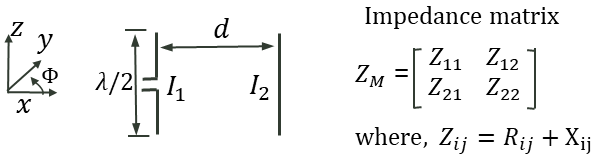}
  \caption{Two element system consisting of half-wave dipoles having currents $I_{1}$ and $I_{2}$, where one is excited and the other is simply shorted.  \label{fig:two_element_diagram}}
\end{figure}

Consider a two element parasitic array in free space consisting of a length $\lambda/2$ antenna and a parasitic element that is also $\lambda/2$ long, as shown in Figure \ref{fig:two_element_diagram}. Without loss of generality, let us also assume the elements are vertically oriented along the $z$ direction. Then, the gain pattern of the two element antenna system (in the $\phi$ plane) relative to the single half-wave dipole (ignoring the loss resistance in the dipoles) can be written as:  


\begin{equation}
G_{d}(\phi) = {\left[\frac{R_{11}}{{R_{11} - \left|\frac{Z_{12}^2}{ Z_{22}}\right|\cos(2\tau_{m} - \tau_{2})}}\right]}^{0.5}\left(1 + \left|\frac{{}Z_{11}}{Z_{22}}\right|\angle{\xi}\right), 
\label{eqn:two_element_gain_phi}
\end{equation}

\noindent where, 

\begin{equation}
\xi = \pi + \tau_{m} - \tau_{2} + \frac{2\pi d }{\lambda}\cos(\phi),
\label{eqn:two_element_xi}
\end{equation}

\begin{equation}
\tau_{m} = \arctan\left[\frac{X_{12}}{R_{12}}\right] \quad \textrm{and} \quad \tau_{2} = \arctan\left[\frac{X_{22}}{ R_{22}}\right],
\end{equation}

\noindent and $Z_{ii}$ and $Z_{ij}$ represent the self-impedance of the $i^{th}$ element and the mutual impedance of the two elements, respectively. Similarly, $R_{ii}$, $X_{ii}$, and $X_{ij}$  are the self-resistance of the $i^{th}$ element, self-reactance of the $i^{th}$ element, and the mutual-reactance of the two elements. A more in-depth derivation can be found in \cite{kraus2002antennas}. 

As the number of elements increases it becomes intractable to derive the gain pattern because the interaction among the various elements becomes prohibitively complex. Based on equation (\ref{eqn:two_element_gain_phi}), the gain pattern of this simple two-element system varies
with the spacing $d$ and the various impedances that vary with the length of the elements (fixed 
at $\lambda / 2$ in the figure).
If, for example, the parasitic element is "de-tuned" by making the reactance component very large (i.e, $X_{22}$) which in turn makes $Z_{22}$ large, then the gain relative to the half-wave dipole given in (\ref{eqn:two_element_gain_phi}) simplifies to 1. This means the effect of the parasitic element becomes minimal and the array loses its gain and directionality. By judiciously selecting the number of parasitic elements and their locations, and length of the various elements, we can obtain better array characteristics in terms of our performance metrics. For multi-element arrays the design space is typically explored empirically or numerically.

A well-known and widely used example of an antenna design based on the idea described above is known as a Yagi-Uda antenna array \cite{thiele1969analysis},\cite{jones1997design}. This parasitic array, which is often designed based on empirical models, uses a linear configuration of parasitic elements, and can provide peak gain as high as 20 dB depending on the total number of elements. A typical design consists of at least one reflector element and multiple director parasitic elements. The reflector element is often made longer than the excited element ensuring that it has an inductive behavior while the director elements are shorter than the excited element which makes them capacitive. This causes the energy to flow predominantly in one direction along the array since the inductive element (the reflector) causes the energy to go back in the direction of the excited element. At the same time,  the capacitive elements (the directors) enhance the propagation in the direction opposite from which it was received. The net effect is a radiation pattern with improved gain in a desired direction.  

There are challenges with this type of design approach. If the elements are in realistic propagation conditions, such as a ground based system, or near a large object or barrier such as a wall, and are allowed to be mobile when the antenna system is deployed, then the impedance relationships may change. This change affects the gain, and therefore the performance, of the parasitic array. We focus on two aspects of the problem in this paper. First, we study the effect of introducing element mobility with ground based UGVs by introducing uncertainties such as exact element location and angular orientation, in Sections \ref{sec:parameteric_study} and \ref{sec:experiment}. Second, we investigate the possibility of re-configuring the geometry (i.e., length and relative locations of the various elements) to improve performance, taking into account realistic ground electromagnetic properties, in Section \ref{sec:optimal_design}. 


\section{Parametric Study Based on Full Wave Simulations}
\label{sec:parameteric_study}
\subsection{Full-Wave Simulation Setup}
In order to investigate the effect of various sources of uncertainties including position and relative orientation of mobile parasitic elements as well as ground effects, we carry out
a full wave simulation that evaluates Maxwell's Equations, over a range of system parameters.
We do this by introducing random errors on the various parameters and performing full wave simulations for each realization of the parameters. This accurately quantifies the error in directivity, beam direction, and side lobe levels while accurately capturing the propagation and coupling effects by solving Maxwell's equations directly using the Finite Difference Time Domain (FDTD) approach with a commercially available solver \cite{EMCUBE}. This helps us determine the most important design variables from which we progress to design variation in the presence of ground scattering, different ground dielectric values, and placement uncertainties associated with node mobility. 

\begin{figure}[t]
  \centering
  \includegraphics[width=0.99\columnwidth]{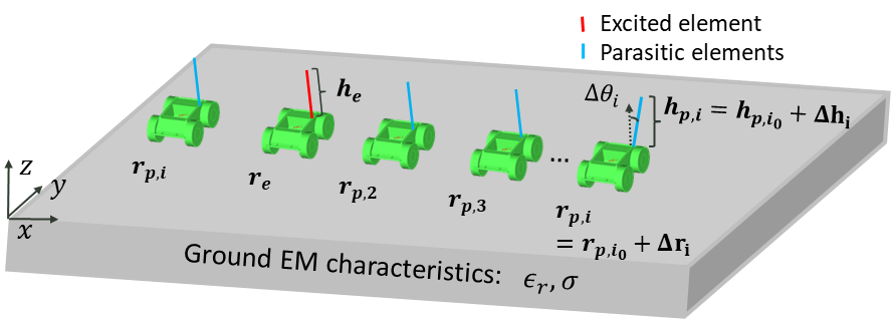}
  \caption{Full wave simulation environment, with the parasitic array near the ground modeled as a homogeneous lossy dielectric slab. The array consists of one excited element and multiple parasitic elements mounted on a UGV model.  \label{fig:simulation_setup}}
\end{figure}

The full wave simulation setup is shown in Figure \ref{fig:simulation_setup}, and simulation parameters are summarized in Table \ref{table:parameters}. We use a homogeneous lossy dielectric medium to model the effects of the ground. A UGV model is also integrated in the simulation on which the excited and parasitic elements are mounted. A single excited element is located at $r_{e}$. The other UGVs are equipped with shorted monopole parasitic elements. We first utilize a baseline Yagi-Uda parasitic array designed in free space and introduce uncertainties in the parameters to quantify the effect on performance. The baseline design has a 5-element linear parasitic array with a reflector element located at $r_{p,1}$, and three director elements located at $r_{p,2}$ to $r_{p,4}$. The initial locations are chosen based on an  empirically optimized design in free space. The resulting baseline design has reflector element $0.25\lambda$ away from the excited element, the distance between the excited element and the first director element is $0.31\lambda$, and the spacing between consecutive director elements is also $0.31\lambda$ \cite{kraus2002antennas}. All the elements in the baseline design are oriented along the $z$ direction. Starting with this baseline design, we include the ground model and perturb the relevant parameters to capture the effects of uncertainties in terms of element location, orientation, as well as the ground electromagnetic characteristics (i.e., the ground dielectric constant and conductivity). 

\begin{table}[t]
\caption{Array Simulation Parameters}
\centering 
\begin{tabular}{|c| c| l|}
\hline\hline                        
Parameter & Values & Notes \\ [0.5ex] 
\hline                 
$f$ & 40MHz     &   Center frequency  \\ 
\hline          
$\lambda$ & 7.5 meters     &   Wavelength at $f=40MHz$  \\ 
\hline 
$\epsilon_{r}$ & [1.5,8]     &   Ground dielectric constant      \\ 
\hline
$\sigma$ & $[10^{-4},10^{-2}]$    & Ground conductivity range \\
\hline
$r_{p,i}$  &  variable    &  Location coordinate of $i^{th}$ parasitic    \\ 
\hline
$r_{e}$  &  variable    &  Location coordinate of excited element    \\ 
\hline
$h_{p,i}$  &  [$0.21\lambda$,$0.26\lambda$]   &  Length of $i^{th}$ parasitic    \\ 
\hline
$h_{e}$  &  $0.26\lambda$   &  Length of excited element  \\ 
\hline
$t_{w}$  &  $4 \times 10^{-2}\lambda$   &  Diameter of all elements   \\ 
\hline
$t_{g}$  &  $6.6 \times 10^{-2}\lambda$   &  Diameter of all elements   \\ 
\hline
$\Delta r_{i}$  &  [$-0.2\lambda$,$0.2\lambda$]  &  Location uncertainity of $i^{th}$ parasitic    \\ 
\hline
$\Delta \theta_{i}$  &  [$-{10^{\circ}},+{10^{\circ}}]$    &  Orientation uncertainity of $i^{th}$ parasitic    \\ 
\hline
\end{tabular}
\label{table:parameters}
\end{table}

\subsection{Position and Orientation Uncertainties}  
The simulations enable accurate study of the ground effects, system parameters, and parameter uncertainties such as robot placement inaccuracy. Robotic accuracy includes both placement and angular orientation, and assumes the robot will estimate its pose with some residual error. Beginning with a good free space design initializes the search space.   

\begin{figure}[t]
  \centering
  \includegraphics[width=0.99\columnwidth]{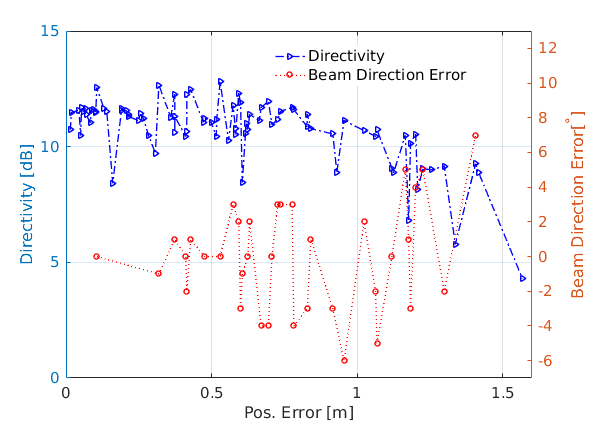}
  \caption{The directivity and error in beam direction for the 5 element parasitic array is shown as a function of position error of the reflector parasitic element. The other director elements are statically positioned (based on a free space design). A uniformly distributed position error in the the parasitic element is introduced. Each data point corresponds to one run of the full wave FDTD simulation. \label{fig:D_vs_ref_pos_error}}
\end{figure}

We first focus on quantifying the effect of position uncertainties on the reflector, which is aligned behind the excited element at a location that ensures the incident and reflected waves add in such a way that the energy is reflected back in the direction of the excited element. In a conventional free space design the reflector alone provides a directional gain of roughly 5dB, and adding parasitics enhances the gain and directivity. We assume a uniformly distributed error in the reflector position, with the other elements ideally place in the baseline design. We choose directivity and beam direction error to measure the performance. The directivity is defined as: 

\begin{equation}
D = \frac{1}{\int_{0}^{2\pi}\int_{0}^{\pi}{|FF_{n}(\theta,\phi)|}^{2}sin(\theta)d_{\theta}d_{\phi}},
\end{equation}

\noindent where $FF_{n}(\theta,\phi)$ is the normalized radiation pattern. The directivity provides a way to measure how directional the radiation is relative to an isotropic antenna. Here, we also define the beam direction as the ($\theta_{i},\phi_{i}$) pair at which the peak of the main beam occurs. Figure \ref{fig:D_vs_ref_pos_error} shows the directivity and the beam direction error as a function of the position error in the reflector node. Errors smaller than 60cm in reflector position, assuming optimal positions for all other nodes, result in a mean directivity of 11.3dB and a mean beam direction error of $1.2^{\circ}$. 

\begin{figure}[h]
  \centering
  \includegraphics[width=0.99\columnwidth]{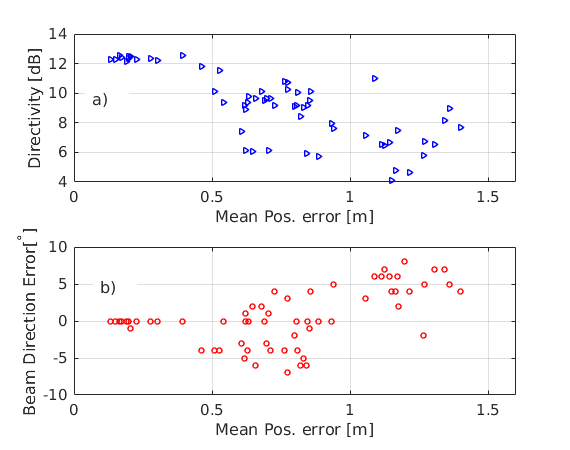}
  \caption{The directivity and error in beam direction for the 5 element parasitic array  as a function of mean position error in the three director elements. Uniformly distributed position errors are introduced in the three director parasitic nodes. Each data point corresponds to one run of the full wave FDTD simulation.  \label{fig:D_vs_directors_pos_error}}
\end{figure}

Next, we investigate the effect of uniformly distributed director element position error. Specifically, position errors up to 1.5m (i.e., $0.2\lambda$) are introduced on each director elements. For each new set of director positions, the full wave simulation is run to compute the directivity and the beam direction and we report mean errors over all realizations simulated. Figure \ref{fig:D_vs_directors_pos_error} shows the directivity and beam direction error versus the mean position error of the three parasitic elements. In this case the reflector is positioned at the optimal location based on the baseline design. Here, the directivity drop is more pronounced than the case where the reflector has position errors (which is shown in Figure \ref{fig:D_vs_ref_pos_error}). This is especially evident as the position error increases beyond 50cm. The errors in beam direction are similar to the first set of simulations where position errors were introduced in the reflector element. 

\begin{figure}[t]
  \centering
  \includegraphics[width=0.97\columnwidth]{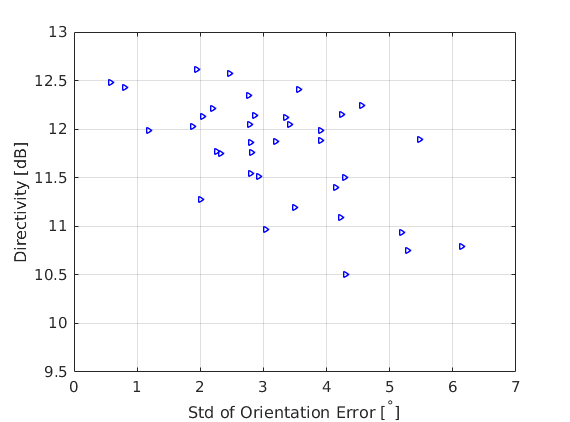}
  \caption{The directivity for the 5 element parasitic array as a function of orientation error in the three director elements. We use the standard deviation of the orientation errors among the three elements as a measure of misalignment. The element misalignment based on this simulation did not result in beam direction errors.     \label{fig:D_vs_directors_orientation_error}}
\end{figure}

We also carried out simulations to quantify the effect of parasitic element angular misalignment. Figure \ref{fig:D_vs_directors_orientation_error} shows the variation in directivity when uniformly distributed random orientation errors are introduced in the three director elements. Element misalignment as high as $\pm6$ degrees causes the directivity to decrease by 2dB. 

One of the main conclusions of the parametric study is that the a certain level of position uncertainty in position of the parasitic elements (i.e., up to 50cm which is equivalent to $0.07\lambda$ at 40MHz) cause relatively small errors both in directivity and beam direction. This level of position accuracy could be achieved with existing solutions. We also note the tolerance to orientation uncertainties as shown in Figure \ref{fig:D_vs_directors_orientation_error}. It should be noted that the required accuracy level will increase if the frequency of operation increases significantly. 


\begin{figure}[t]
  \centering
  \includegraphics[width=0.97\columnwidth]{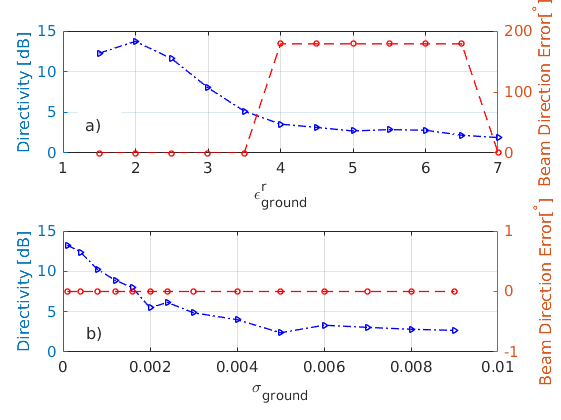}
  \caption{The directivity and error in beam direction for the 5 element parasitic array are shown as a function of the real part (a) and conductivity (b) of the ground electromagnetic characteristics. It is seen that as dielectric constant and conductivity increase, the directive gain decreases. This is because the free space design is no longer valid and a design with a different element configuration with mobility could perform better. It should be noted that variation in conductivity alone does not affect the beam direction. \label{fig:D_vs_grd_diel}}
\end{figure}

\subsection{Effects of Ground Electromagnetic Characteristics}
Another important parameter, especially for UGV mounted antennas, is the effect of the ground electromagnetic characteristics. Most existing designs are carried out in isolation or free space scenarios. Since parasitic arrays such as Yagi-Uda are travelling wave antennas, the effect of nearby objects can be significant. Here, we study the effect of the ground for a parasitic array and propose a design strategy for an optimal design depending on the specific ground characteristics, which we model as a homogeneous lossy dielectric medium with an effective complex dielectric constant. We performed two sets of simulations where we vary relative dielectric constant and conductivity of the ground, which represent the delay and loss, respectively. Figure \ref{fig:D_vs_grd_diel}a shows the directivity and beam direction error as a function of the real part of the ground dielectric constant. In this specific simulation, as the real part of the dielectric constant increases above 3.5, the directive gain effectively goes away and the main beam points in an unintended direction. We also studied the performance of the parasitic array as a function of the conductivity of the ground. The results are shown in Figure \ref{fig:D_vs_grd_diel}b. As the conductivity increases, the gain of the array decreases rapidly. As the dielectric constant and conductivity increase the free space design becomes sub-optimal. This is a primary motivation for leveraging mobile parasitic nodes that enable  performance in varying  propagation conditions. We show an example design for a higher dielectric ground case in a later section.

\section{UGV Experiment}
\label{sec:experiment}
\begin{figure}[t]
  \centering
  \includegraphics[width=0.9\columnwidth]{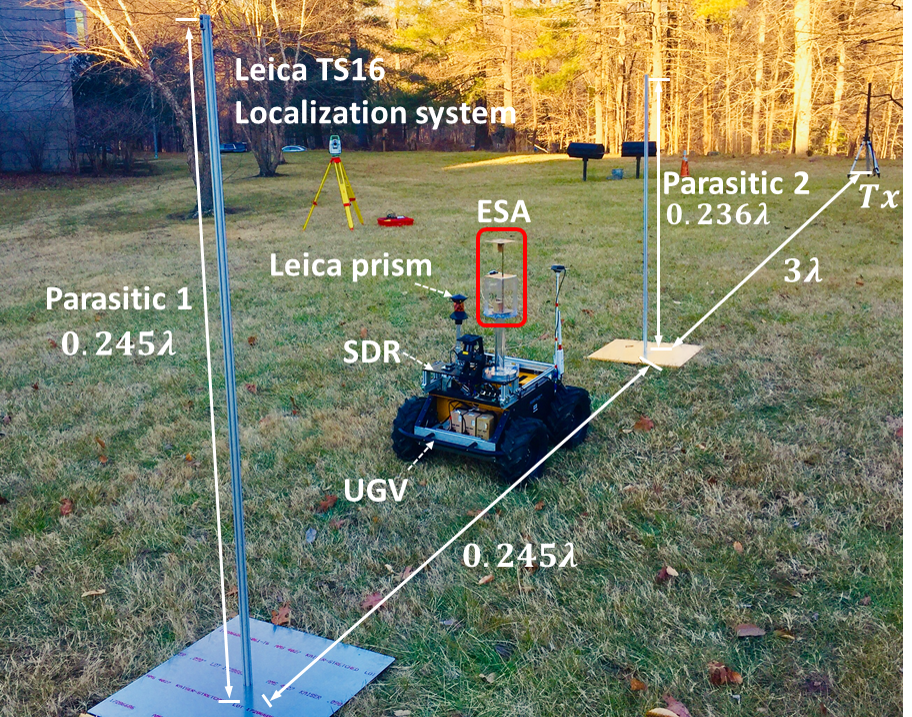}
  \caption{ Experiment setup for a 3-element parasitic array: In these tests parasitic element 1 is used as a reflector, parasitic element 2 is used as a director, and both elements are shorted aluminum monopoles with slight difference in height. In our experiment, we drive the $\lambda / 25$ ESA equipped robot to different points between the parasitic elements. The transmitter (Tx) is a miniature short dipole ($\lambda/6$) antenna.  The size and spacing of the antenna elements are overlayed on the image.  We use the Leica TS16 to accurately measure the location of the antennas.
  \label{fig:experimental_setup}.}
\end{figure}

We describe an experiment using a 3 element parasitic array with one mobile element, shown in Figure \ref{fig:experimental_setup}. A compact UGV is equipped with an Ettus N210 software defined radio (SDR) and an electrically small antenna (i.e., $\lambda/25$ operating at 40MHz center frequency). The UGV mounted ESA is highly efficient and provides an omni-directional radiation pattern \cite{Choi2015electrically}.

Note that conventional parasitic arrays often use half-wave dipoles. Our setup also includes two static aluminum parasitic elements. These parasitic elements are shorted monopoles with a slight difference in height to achieve the inductive and capacitive behavior described in Section \ref{sec:basics}. This 3-element system is tested as a receive parasitic array. We also use a short dipole antenna ($\lambda/6$ length) that transmits a tone at 40MHz (here $\lambda = 7.5m$) using a signal generator with the the transmit power set at $10mW$. The receive radio on the robot collects $5\times10^{5}$ IQ samples at a sampling rate of $2\times10^6samples/sec$ for each location of the robot. For all configurations, we measure the positions of the elements with high precision  using a Leica Viva TS16 localization system \cite{ts16}. We initially configure all four elements in a linear fashion. We then move the UGV to different locations between the two parasitic elements. For each new location, we collect the $I$ and $Q$ samples of the received signal which are then used to compute the received signal strength. The position of the ESA mounted on the mobile agent with respect to the reflector and director is the independent variable in this experiment.  In each trial we position the mobile agent, record the  $I$ and $Q$ samples in the presence of the parasitic elements, and then record the same data without the parasitic elements. The data collected without the parasitic elements serves as the control. 

\begin{figure}[h]
  \centering
  \includegraphics[width=0.99\columnwidth]{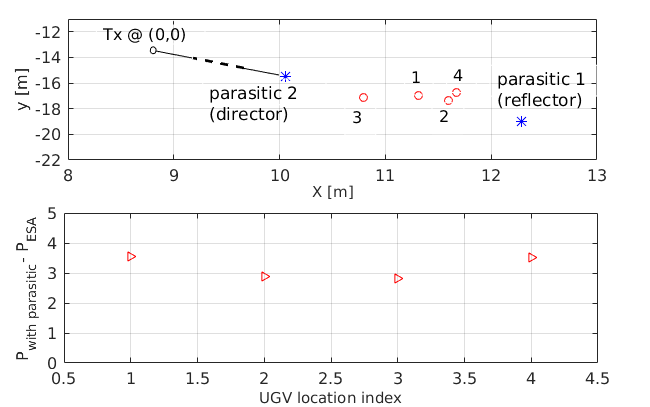}
  \caption{The relative gain of the parasitic array compared to the ESA is shown in dB. The element configuration is a conventional free space design. The ESA is used as the experimental director element. 
  \label{fig:experiment_result}}
\end{figure}

The results of this experiment are shown in Figure \ref{fig:experiment_result}. The configurations tested here are based on the free space conventional design described previously, that is typically used with an elevated antenna which reduces scattering from the environment. The top subplot in Figure \ref{fig:experiment_result} shows locations of the various elements measured using the TS16 system. In the bottom plot, we show the relative gain of the parasitic array. For each configuration we calculate the relative gain by taking the difference between the received power measured with and without the parasitic elements. The mean gain for this experiment was 3.2dB. In the free space scenario, the 3 element array provides 6.5dB gain. We also conjecture that in addition to the parasitic element locations and length, the use of the ESA in the measurements can also contributed to the reduced gain versus free space operation. Ideally the parasitic elements will be designed to maximize their coupling with the ESA. The parasitic elements used here are optimally sized for a monopole antenna operating in free space. To optimize the overall design for the near ground case studied here, next we apply a genetic algorithm to find placements that significantly enhance performance.



\begin{figure}[t]
  \centering
  \includegraphics[width=1\columnwidth]{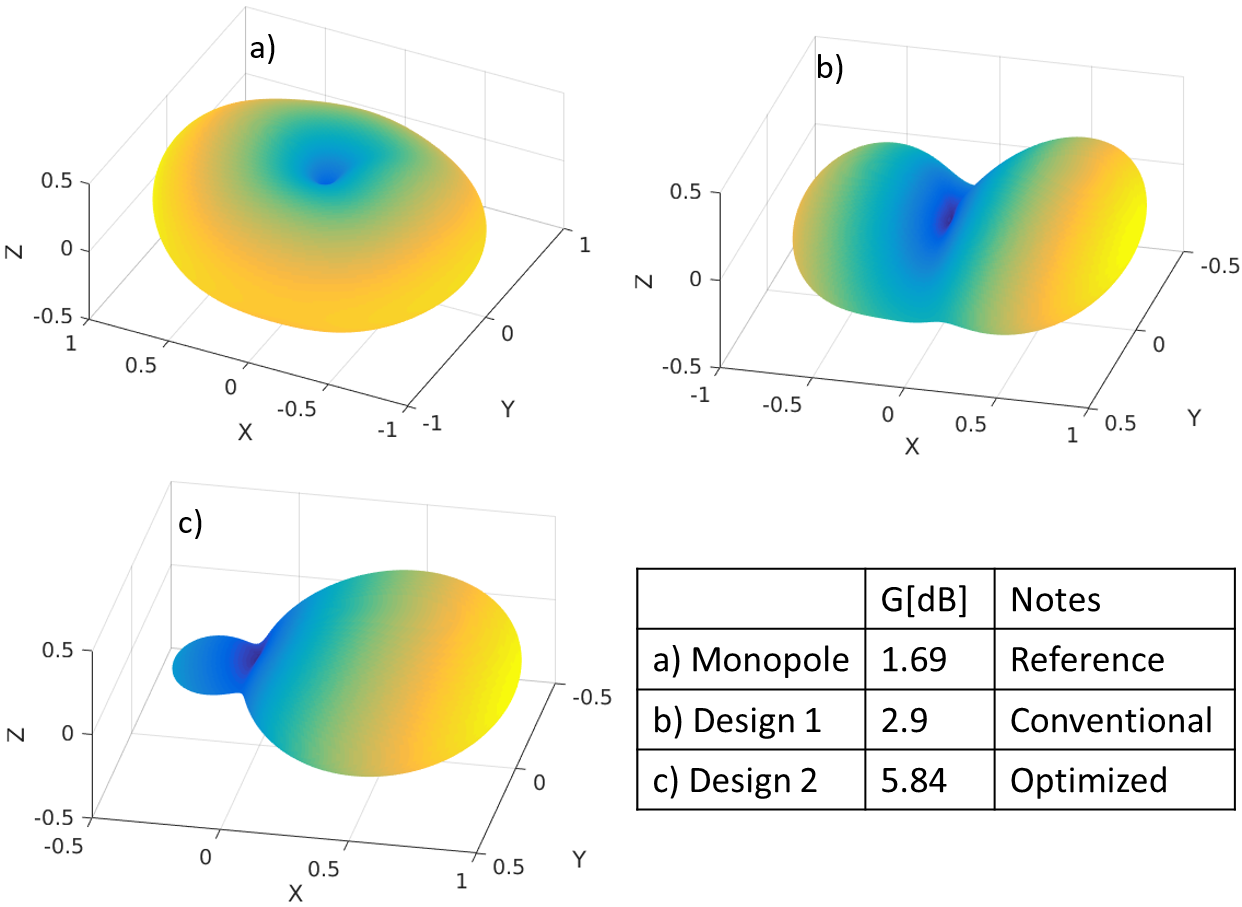}
  \caption{3D radiation patterns for three cases on a ground having dielectric constant and conductivity of concrete.  a) Reference monopole antenna mounted on the UGV.  b) Two parasitic elements are added to the monopole using the conventional free space configuration. c) Optimized configuration using a genetic algorithm.  The optimal design incorporates the ground effects to yield significantly improved performance,}
  \label{fig:optimization_result}
\end{figure}

\section{Optimizing Array Design in the Presence of Ground Dielectric}
\label{sec:optimal_design}
The full wave parametric study and experiments described above indicate that the parasitic array performed reasonably well even in the presence of uncertainties in linear position and angular orientation. The level of tolerable position uncertainties is a strong function of center frequency or corresponding wavelength.  The parasitic design is a critical function of the element positioning with respect to wavelength, so that lower frequency (longer wavelength) implies less sensitivity to a fixed position error distance.  However, we also want to characterize the array performance in the presence of a ground dielectric that deviates from free space. As shown in Figure \ref{fig:D_vs_grd_diel}, higher dielectric constant and conductivity of the ground can significantly degrade the performance of the parasitic array if the conventional free space design is used. Both the directive gain and the beam direction are significantly affected. We explore the design space by employing a genetic algorithm optimization in conjunction with the full wave FDTD solver, to find system parameters that account for the ground electromagnetic properties.  The system parameters include the placement and heights of the parasitic elements.


Consider the 3 element linear setup along the $X$ axis (see Figure \ref{fig:simulation_setup}). The ground is modeled as a homogeneous medium made of concrete; a lossy medium with a dielectric constant of $\epsilon_r = 4.5$ and conductivity $\sigma = 0.01$. The length of each parasitic element and the spacing between the excited element and the parasitic elements are the parameters to be optimized. The objective functions we choose are: 

\begin{equation}
|FF_{D} - g_{d}| <= \epsilon_g,  
\label{eqn:objective_gain}
\end{equation}

\begin{equation}
|FF_{MB}^{Az} - 0| <= \epsilon_z,  
\label{eqn:objective_gain2}
\end{equation}

\begin{equation}
|FF_{MB}^{El} - 90| <= \epsilon_l, 
\label{eqn:objective_gain3}
\end{equation}

\noindent where $FF_{D}$, $|FF_{MB}^{Az}$, and $|FF_{MB}^{El}$ are the peak gain, and main beam directions in the azimuth and elevation planes, respectively. $g_{d}$ is the desired peak gain of the array. Note that other objectives could also be utilized, including reducing the side lobe levels and beam width. The error tolerance for optimization  convergence is denoted by $\epsilon_i$. It should be noted that we limit the search space for the position parameters to along the axis of the array for the optimization results presented here. This means we are optimizing the spacing between the excited element and the two parasitic elements as well as the length of the two parasitic elements. The resulting values for the spacing between the director and the excited element was $0.16\lambda$ which is a 48.4\% reduction compared to the conventional free space design. The optimal spacing between the reflector element and the excited element was $0.14\lambda$ which is a 44\% decrease compared to the conventional free space design. The optimal length of the the director decreased by 8.8\% and the length of the reflector increased only slightly by 0.3\%. Figure \ref{fig:optimization_result} shows the radiation patterns of the excited antenna without parasitic elements (i.e., the reference monopole), the parasitic array with conventional free space configuration, and the optimized configuration. It can be seen in Figure \ref{fig:optimization_result}(c) that the optimized configuration provides a significantly improved directional radiation in the presence of the ground compared to the conventional design which is shown in Figure \ref{fig:optimization_result}(b).

\section{Conclusion}
Low frequency communications with electrically small antennas enables robots to network in complex environments due to the enhanced propagation; long wavelengths penetrate structures and obstacles, and small antennas are consistent with robotic deployment.  We considered the use of a multi-element antenna array to manipulate the radiation pattern and form directional beams.  Using parasitic elements and only a single active element simplifies a multi-robot deployment while providing good beam patterns.  
    
 We performed full wave FDTD simulations of Maxwell's equations in order to a accurately determine a favorable configuration of parasitic elements and estimate the array radiation pattern.  This approach relies critically on the cross-element coupling that is sensitive to the positioning relative to the signal phase. Then, we tested our simulation results with an outdoor experiment employing a mobile parasitic element.  Finally, we demonstrated the ability to optimize array configuration in the presence of a ground dielectric. Performance of a near ground UGV deployment depends critically on the electromagnetic properties of the ground and the ability to physically manipulate the array enables good array patterns.  Using this knowledge we show it is possible to place a mobile agent in relation to parasitic elements in order to increase directivity.
  
 This research is only a first step towards multi-robot radiation pattern manipulation.  We are interested in varying array shapes, antenna types and miniaturization (i.e., the excited and the parasitic elements) and number of nodes in different environments.  We also plan to develop algorithms which guide a robot or team of robots to modify their positions in order to manipulate the radiation patterns.    

\bibliographystyle{IEEEtran}
\bibliography{bib.bib}

\begin{thebibliography}{10}
\providecommand{\url}[1]{#1}
\csname url@rmstyle\endcsname
\providecommand{\newblock}{\relax}
\providecommand{\bibinfo}[2]{#2}
\providecommand\BIBentrySTDinterwordspacing{\spaceskip=0pt\relax}
\providecommand\BIBentryALTinterwordstretchfactor{4}
\providecommand\BIBentryALTinterwordspacing{\spaceskip=\fontdimen2\font plus
\BIBentryALTinterwordstretchfactor\fontdimen3\font minus
  \fontdimen4\font\relax}
\providecommand\BIBforeignlanguage[2]{{%
\expandafter\ifx\csname l@#1\endcsname\relax
\typeout{** WARNING: IEEEtran.bst: No hyphenation pattern has been}%
\typeout{** loaded for the language `#1'. Using the pattern for}%
\typeout{** the default language instead.}%
\else
\language=\csname l@#1\endcsname
\fi
#2}}

\bibitem{Choi2015electrically}
J.~Choi, F.~T. Dagefu, B.~M. Sadler, and K.~Sarabandi, ``Electrically small
  folded dipole antenna for hf and low-vhf bands,'' \emph{IEEE Antennas and
  Wireless Propagation Letters}, vol.~15, pp. 718--721, 2016.

\bibitem{oh2013extremely}
J.~Oh, J.~Choi, F.~T. Dagefu, and K.~Sarabandi, ``Extremely small two-element
  monopole antenna for hf band applications,'' \emph{Antennas and Propagation,
  IEEE Transactions on}, vol.~61, no.~6, pp. 2991--2999, 2013.

\bibitem{williams1993virtual}
D.~Williams and O.~Khatib, ``The virtual linkage: A model for internal forces
  in multi-grasp manipulation,'' in \emph{Robotics and Automation, 1993.
  Proceedings., 1993 IEEE International Conference on}.\hskip 1em plus 0.5em
  minus 0.4em\relax IEEE, 1993, pp. 1025--1030.

\bibitem{song2002potential}
P.~Song and V.~Kumar, ``A potential field based approach to multi-robot
  manipulation,'' in \emph{Robotics and Automation, 2002. Proceedings. ICRA'02.
  IEEE International Conference on}, vol.~2.\hskip 1em plus 0.5em minus
  0.4em\relax IEEE, 2002, pp. 1217--1222.

\bibitem{spletzer2001cooperative}
J.~Spletzer, A.~K. Das, R.~Fierro, C.~J. Taylor, V.~Kumar, and J.~P. Ostrowski,
  ``Cooperative localization and control for multi-robot manipulation,'' in
  \emph{Intelligent Robots and Systems, 2001. Proceedings. 2001 IEEE/RSJ
  International Conference on}, vol.~2.\hskip 1em plus 0.5em minus 0.4em\relax
  IEEE, 2001, pp. 631--636.

\bibitem{mainprice2013human}
J.~Mainprice and D.~Berenson, ``Human-robot collaborative manipulation planning
  using early prediction of human motion,'' in \emph{Intelligent Robots and
  Systems (IROS), 2013 IEEE/RSJ International Conference on}.\hskip 1em plus
  0.5em minus 0.4em\relax IEEE, 2013, pp. 299--306.

\bibitem{fink2013robust}
J.~Fink, A.~Ribeiro, and V.~Kumar, ``Robust control of mobility and
  communications in autonomous robot teams,'' \emph{IEEE Access}, vol.~1, pp.
  290--309, 2013.

\bibitem{lui2009note}
H.-S. Lui, H.~T. Hui, and M.~S. Leong, ``A note on the mutual-coupling problems
  in transmitting and receiving antenna arrays,'' \emph{IEEE Antennas \&
  Propagation Magazine}, vol.~51, no.~5, pp. 171--176, 2009.

\bibitem{huang2006mutual}
Z.~Huang, C.~A. Balanis, and C.~R. Birtcher, ``Mutual coupling compensation in
  ucas: Simulations and experiment,'' \emph{IEEE Transactions on Antennas and
  Propagation}, vol.~54, no.~11, pp. 3082--3086, 2006.

\bibitem{liao2012adaptive}
B.~Liao and S.-C. Chan, ``Adaptive beamforming for uniform linear arrays with
  unknown mutual coupling,'' \emph{IEEE Antennas and Wireless Propagation
  Letters}, vol.~11, pp. 464--467, 2012.

\bibitem{kraus2002antennas}
J.~D. Kraus, R.~J. Marhefka, and A.~S. Khan, \emph{Antennas for all
  applications}.\hskip 1em plus 0.5em minus 0.4em\relax McGraw-Hill, 2002.

\bibitem{graefenstein2009wireless}
J.~Graefenstein, A.~Albert, P.~Biber, and A.~Schilling, ``Wireless node
  localization based on rssi using a rotating antenna on a mobile robot,'' in
  \emph{Positioning, Navigation and Communication, 2009. WPNC 2009. 6th
  Workshop on}.\hskip 1em plus 0.5em minus 0.4em\relax IEEE, 2009, pp.
  253--259.

\bibitem{cliff2015online}
O.~M. Cliff, R.~Fitch, S.~Sukkarieh, D.~Saunders, and R.~Heinsohn, ``Online
  localization of radio-tagged wildlife with an autonomous aerial robot
  system.'' in \emph{Robotics: Science and Systems}, 2015.

\bibitem{zhang2004pattern}
S.~Zhang, G.~Huff, J.~Feng, and J.~Bernhard, ``A pattern reconfigurable
  microstrip parasitic array,'' \emph{IEEE Transactions on Antennas and
  Propagation}, vol.~52, no.~10, pp. 2773--2776, 2004.

\bibitem{petit2006mems}
L.~Petit, L.~Dussopt, and J.-M. Laheurte, ``Mems-switched parasitic-antenna
  array for radiation pattern diversity,'' \emph{IEEE Transactions on Antennas
  and Propagation}, vol.~54, no.~9, pp. 2624--2631, 2006.

\bibitem{thiele1969analysis}
G.~Thiele, ``Analysis of yagi-uda-type antennas,'' \emph{IEEE Transactions on
  Antennas and Propagation}, vol.~17, no.~1, pp. 24--31, 1969.

\bibitem{jones1997design}
E.~A. Jones and W.~T. Joines, ``Design of yagi-uda antennas using genetic
  algorithms,'' \emph{IEEE Transactions on Antennas and Propagation}, vol.~45,
  no.~9, pp. 1386--1392, 1997.

\bibitem{EMCUBE}
\BIBentryALTinterwordspacing
Emag technologies inc. [Online]. Available: \url{https://www.emagtech.com}
\BIBentrySTDinterwordspacing

\bibitem{ts16}
\BIBentryALTinterwordspacing
Leica. (2019) Leica viva ts16 - world’s first self-learning total station.
  [Online]. Available:
  \url{https://leica-geosystems.com/en-US/products/total-stations/robotic-total-stations/leica-viva-ts16}
\BIBentrySTDinterwordspacing

\end{thebibliography}

\end{document}